\documentclass[%
reprint,
superscriptaddress,
 amsmath,amssymb,
 aps,
floatfix,
]{revtex4-1}

\usepackage{hyperref}
\hypersetup{
  colorlinks=true,
  linkcolor=blue,
  urlcolor=cyan,
}

\usepackage{physics} 
\usepackage{amssymb} 
\usepackage{bm} 
\usepackage{braket} 
\usepackage{mathtools} 
\usepackage{graphicx} 
\graphicspath{{./figures/}} 
\usepackage{float}

\newcommand{\JILA}{JILA, National Institute of Standards and Technology and Department of Physics, University of Colorado, Boulder, CO, 80309, USA}
\newcommand{\CTQM}{Center for Theory of Quantum Matter, University of Colorado, Boulder, CO, 80309, USA}
\newcommand{\PURDUE}{Department of Physics and Astronomy, Purdue University, West Lafayette, IN, 47907, USA}
\newcommand{\PURDUESecond}{
Purdue Quantum Science and Engineering Institute, Purdue University, West Lafayette, IN, 47907, USA}

\begin{document}

\title{Understanding chemical reactions in a quantum degenerate gas of polar molecules via complex formation  }

\author{Peiru He}
\affiliation{\JILA}
\affiliation{\CTQM}
\author{Thomas Bilitewski}
\affiliation{\JILA}
\affiliation{\CTQM}
\author{Chris H. Greene}
\affiliation{\PURDUE}
\affiliation{\PURDUESecond}
\author{Ana Maria Rey}
\affiliation{\JILA}
\affiliation{\CTQM}

\date{\today}

\begin{abstract}
A recent experiment \cite{de2019degenerate} reported for the first time the preparation of a Fermi degenerate gas of polar molecules and observed a suppression of their chemical reaction rate compared to the one expected from a purely classical treatment. While it was hypothesized that the suppression in the ultracold regime had its roots in the Fermi statistics of the molecules, this argument  is inconsistent with  the fact that the Fermi pressure should set a lower bound for the chemical reaction rate. 
Here we develop a simple model of chemical reactions that occur via the formation and decay of molecular complexes.
We indeed find that pure two-body molecule losses are unable to explain the observed suppression. Instead we extend our description beyond two-body physics by including  effective complex-molecule interactions possible emerging from    many-body and effective medium effects at finite densities and in the presence of trapping light. 
Although our effective model is able to quantitatively reproduce recent  experimental observations,  a detailed understanding  of the actual physical mechanism  responsible for these higher-order interaction processes is still pending. \end{abstract}

\maketitle

\textit{Introduction.}
Polar molecular gases,
offering tunable long-range interactions and a large set of internal degrees of freedom, 
are an ideal platform to explore a wide range of many-body phenomena that are difficult to access in atomic systems.
The prerequisite for many of these explorations is the preparation of quantum degenerate samples,
which has been one of the most challenging goals in molecular physics over past decades ~\cite{carr2009cold,moses2017new,moses2015creation,anderegg2018laser,de2019degenerate}.
Major challenges arise due to the complex molecular internal structure and the rapid loss caused by chemical reactions which prevent the application of standard cooling techniques for atoms \cite{carr2009cold,ospelkaus2010quantum}.

\begin{figure}[hbt!]
\centering
\includegraphics[width=0.4\textwidth]{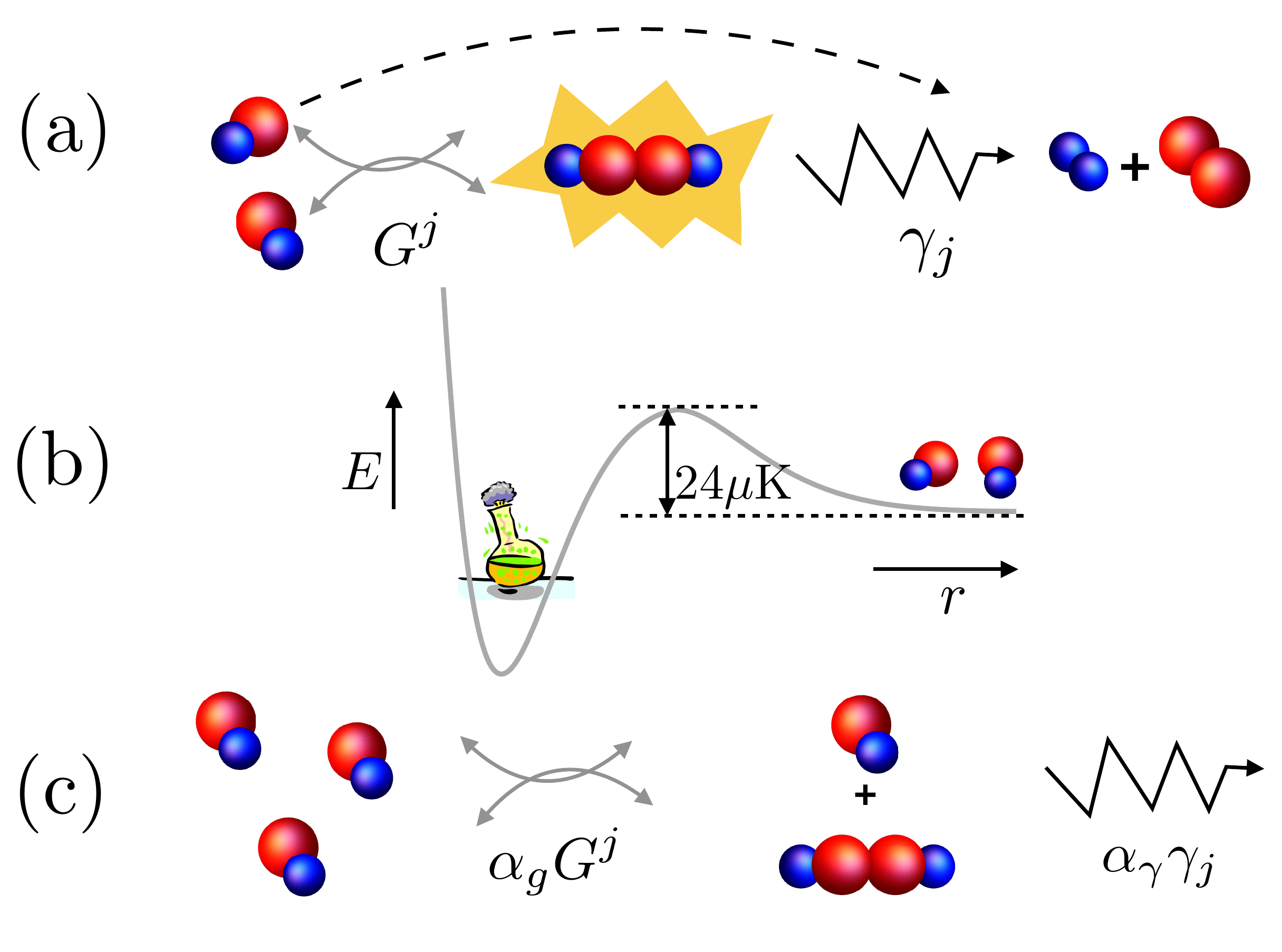}
\caption{Schematics of the reactive collision processes.
(a) Two $\text{KRb}$ molecules coherently collide in the $p$-wave channel with rate $G^j$ to form an intermediate complex $\text{K}_2\text{Rb}_2$, 
which subsequently decays to the reaction products $\text{K}_2$ and $\text{Rb}_2$ at a rate $\gamma_j$. If the complex decay rate $\gamma_j$ is the fastest process, as for  KRb molecules, the complex can be adiabatically eliminated,  giving rise to an effective two-body decay.  (b) This recovers the  standard picture of  direct chemical reactions via  p-wave inelastic  collisions, where chemical reactions  occur with unit probability at short-range inside  the centrifugal barrier. (c) Additional (in)elastic complex-molecule collisions with rate $\alpha_g G^j (\alpha_{\gamma}\gamma_j)$ effectively generate three-body molecule processes which can suppress the two-body molecule decay rate. 
\label{fig:schematic_plot}}
\end{figure}
The use of spin-polarized fermionic molecules facilitated experimental efforts to reduce the undesirable chemical reactions as in these systems the collisions are dominated by $p$-wave scattering. In this case, according to the Bethe-Wigner threshold law \cite{bethe1935theory,wigner1948behavior,sadeghpour2000collisions}, chemical reactions are partially  suppressed by the centrifugal barrier resulting in a loss rate that scales linearly with temperature $T$. A quantitative analysis using a multichannel quantum defect theory (MQDT) \cite{greene1982general,idziaszek2010universal} captured this behavior with a universal decay constant which well explained the experimentally observed decay rate in a gas of $\text{KRb}$ molecules prepared in the classical regime ($T>0.5T_F$, with $T_F$ the Fermi temperature)~\cite{de2019degenerate, ospelkaus2010quantum}. However,  the Bethe-Wigner threshold law  has been shown to fail in a recent experiment~\cite{de2019degenerate} which prepared for the first time a quantum degenerate gas of $\text{KRb}$ molecules in a $3$D dipole trap reaching temperatures below $0.3 T_F$.
Deep in the quantum degenerate limit ($T<0.5T_F$), a significant suppression of the loss rate compared to the one predicted by the MQDT theory was observed and conjectured to be a consequence of the underlying Fermi statistics.
Yet, this explanation is inconsistent with the naive expectation that, as the temperature vanishes, the Fermi pressure sets a lower bound for the $p$-wave reaction rate, which would instead lead to a rate higher than the one predicted by purely classical arguments. The observed suppression therefore requires an explanation more profound than just Fermi statistics.

Recent experiments \cite{hu2019direct,liu2020steering} moreover  revealed that even in reactive molecules such as $\text{KRb}$, chemical reactions occur via the formation of a transient complex whose properties may affect the collision outcome. These observations therefore have opened the possibility of richer chemical reaction processes\cite{croft2020unified}.





Here we provide a possible explanation of the observed chemical reaction suppression at ultracold temperatures by developing a theoretical many-body framework that accounts for the formation of molecular complexes. The large decay rate of the complex ~\cite{hu2019direct,liu2020steering} allows us to adiabatically eliminate the complex, and obtain  an effective two-body decay of the molecules which  recovers the standard description of KRb chemical reactions. We analytically solve the rate equations, accounting for both heating effects and quantum Fermi statistics. We obtain a  decay rate that is  in agreement with the classical Bethe-Wigner threshold laws above quantum degeneracy, and   also valid in the ultracold quantum  regime. However, this model  fails to capture the experimental observations in the quantum degenerate regime.
We therefore turn to an effective description, modelling beyond two-body physics by  including effective elastic and inelastic complex-molecule interactions possibly emerging from many-body and effective medium effects at finite densities and in the presence of trapping light, which can generate  a loss  suppression mechanism alike to the one observed in  the experiment.


\textit{The model.}
We begin by deriving a framework including an intermediate complex, whose existence has recently been experimentally demonstrated \cite{hu2019direct,liu2020steering}, formed via the collision of two molecules as illustrated in Fig.~\ref{fig:schematic_plot}(a), which recovers standard chemical reaction rate equations.

We consider $N$ fermionic molecules, with mass $m$ confined by an  external potential $V({\bm r})$, which for simplicity we first set  to be a simple square well that defines  a confinement volume $V$. In this system  momentum  $\hbar \bf{k}$ is a good quantum number. 
For molecules prepared in a single internal quantum state, $p$-wave scattering dominates the collisions at ultracold temperatures due to Fermi statistics, which is, thus, the only partial wave we include.
Assuming there are multiple channels  to form a complex (each denoted by j) the collision  processes can be modeled by a simplified  master equation 
\begin{eqnarray}
&&\frac{d\hat{\rho}}{dt}=\frac{i}{\hbar}[\hat{H},\, \hat{\rho}]+\mathcal{L}(\hat\rho),
\,\quad \hat{H}=\hat{H}_{\rm {single}} +\hat{H}_{\rm {int}}  \\
&&\hat{H}_{\rm {single}} = \sum_{j,{\bm k}}E^b_{j,{\bm k}}\hat{b}^{\dagger}_{j,{\bm k}}\hat{b}_{j,{\bm k}}+ \sum_{{\bm k}}E^c_{{\bm k}}\hat{c}^{\dagger}_{{\bm k}}\hat{c}_{{\bm k}}
\\&&\hat{H}_{\rm {int}} =\sum_{j, {\bm k}, {\bm k}'}\frac{\hbar g_j}{\sqrt{V}}|{\bm k}-{\bm k}'|
 \big(
\hat{b}^{\dagger}_{j,{\bm k}+{\bm k}'}\hat{c}_{{\bm k}}\hat{c}_{{\bm k}'}
+\text{h.c.}
\big),
\\&&
\mathcal{L}(\hat{\rho})=\sum_{j,{\bm k}}\gamma_j
\mathcal{L}[\hat{b}_{j,{\bm k}}]\,\hat{\rho},
\label{eq:TwoChannel}
\end{eqnarray}
where $\hat{c}^{\dagger}_{{\bm k}}(\hat{c}_{{\bm k}})$ is a fermionic creation(annihilation) operator of a molecule with momentum
$\hbar {\bm k}$, $\hat{b}^{\dagger}_{j,{\bm k}}(\hat{b}_{j,{\bm k}})$ is a bosonic creation(annihilation) operator of a complex formed via channel $j$, $E^c_{j,{\bm k}}=\hbar^2 {\bm k}^2/(2m)$ and $E^b_{j,{\bm k}}=\hbar^2 {\bm k}^2/(4m)+E_j$ the  single-particle energies of the molecules and complexes respectively, with  $E_j$  the binding energy of a complex. The parameter $g_j$ sets the complex-molecule collision strength and $\gamma_j$ is the complex decay rate. 
The Lindblad term,
$\mathcal{L}[\hat{O}]\hat{\rho}=\hat{O}^{\dagger}\hat{\rho}\,\hat{O}-\frac{1}{2}(\hat{\rho}\,\hat{O}^{\dagger}\hat{O}+\hat{O}^{\dagger}\hat{O}\hat{\rho})$, describes the action of an operator $\hat{O}$ on the density matrix $\hat{\rho}$ of the complex-molecule many-body system.  

From the master equation one can obtain equations of motion of  the relevant observables. 
Since for the problem of interest  the  initial state has  zero coherence terms $\langle\hat{c}^{\dagger}_{{\bm k}}\hat{c}_{{\bm k}'}\rangle=0$ and $\langle\hat{c}_{{\bm k}}\hat{c}_{{\bm k}'}\rangle=0$  for ${\bm k}\neq {\bm k}'$, these terms can be neglected during the dynamics giving rise to the following equations:
\begin{eqnarray} 
&& \frac{d \langle \hat{n}_{{\bm k}}\rangle}{dt}
=
\sum_{j,{\bm k}'} 2 G^j_{{\bm k},{\bm k}'}\,
\text{Im}[\langle\hat{b}^{\dagger}_{j,{\bm k}+{\bm k}'} \, \hat{c}_{{\bm k}} \hat{c}_{{\bm k}'}\rangle]
\label{eq:aNField2} \\
&&\frac{d \langle\hat{b}_{j,{\bm k}+{\bm k}'}^{\dagger}\hat{c}_{{\bm k}} \hat{c}_{{\bm k}'}\rangle}{dt} =
i \Big(\frac{\hbar|{\bm k}-{\bm k}'|^2}{4m} - E_j/\hbar +i\gamma_j\Big)
\langle\hat{b}_{j,{\bm k}+{\bm k}'}^{\dagger}\hat{c}_{{\bm k}} \hat{c}_{{\bm k}'}\rangle \notag\\
&&\qquad \qquad\qquad \qquad +i G^j_{{\bm k},{\bm k}'}
\Big(
\langle \hat{n}^b_{j,{\bm k}+{\bm k}'}\rangle
-2\langle\hat{n}_{{\bm k}}\hat{n}_{{\bm k}'}\rangle
\Big)
\label{eq:baFieldc}\\
&&\frac{d \langle\hat{n}^b_{j,{\bm k}}\rangle}{dt} =
-2\gamma_j\langle\hat{n}^b_{j,{\bm k}}\rangle
-G^j_{{\bm k},{\bm k}-{\bm k}'} \,\text{Im}[\langle\hat{b}_{j,{\bm k}}^{\dagger}\hat{c}_{{\bm k}'}\hat{c}_{{\bm k}-{\bm k}'}\rangle]
\label{eq:bbField2}
\end{eqnarray} with  $\hat{n}_{{\bm k}}=\hat{c}^{\dagger}_{{\bm k}}\hat{c}_{{\bm k}}$, $\hat{n}^b_{j,{\bm k}}=\hat{b}_{j,{\bm k}}^{\dagger}\hat{b}_{j,{\bm k}} $ and $G^j_{{\bm k},{\bm k}'} =\frac{2g_j}{\sqrt{V}}|{\bm k}-{\bm k}'|$. The mean complex decay rate $\overline{\gamma}$ has been measured to be $ 2\pi \times 4$MHz in free space and even larger in the presence of trapping light \cite{liu2020steering}. Because this rate is  much larger than any other energy scales of the molecular gas \cite{de2019degenerate}, we can adiabatically eliminate the complexes, and set to zero the  left hand side of Eq.~(\ref{eq:baFieldc}) and
$\langle\hat{n}^b_{j,{\bm k}}\rangle$. The  complex-molecule coherence term then obeys: 
$\langle\hat{b}_{j,{\bm k}+{\bm k}'}^{\dagger}\hat{c}_{{\bm k}} \hat{c}_{{\bm k}'}\rangle\approx
-i \frac{2G^j_{{\bm k},{\bm k}'}}{(\gamma_j-iE_j/\hbar)}
\langle \hat{n}_{{\bm k}}\hat{n}_{{\bm k}'}\rangle\approx-i  \frac{2 G^j_{{\bm k},{\bm k}'}}{(\gamma_j-iE_j/\hbar)}
\langle\hat{n}_{{\bm k}}\rangle\langle \hat{n}_{{\bm k}'}\rangle
$~\cite{supp}. 
Using this in Eqs. (5) and (7)  recovers the standard equations that  describe  direct chemical reactions, if we  identify the $p$-wave collision parameters in terms of the real and imaginary parts of the scattering volume $b_{\text{im},\text{re}}^3$ as follows:  $g_{\text{im}}\equiv 3 \pi \hbar b_{\text{im}}^3/m= \sum_j 4g_j^2 \gamma_j/\big(\gamma_j^2+(E_j/\hbar)^2\big)$  and $g_{\text{re}}\equiv  3 \pi \hbar b_{\text{re}}^3/m= \sum_j 4g_j^2 (E_j/\hbar)/\big(\gamma_j^2+(E_j/\hbar)^2\big)$ (see \cite{supp}). The real part describes elastic collisions that thermalize the system, and the imaginary part gives rise to the reactive collision rate \cite{julienne2009ultracold} as illustrated  in Fig.~\ref{fig:schematic_plot}(b). We observe that in the limit of a large decay rate $\gamma_j \gg g_j$, we are in the quantum Zeno regime \cite{Misra1977,Itano1990,Zhu2013} where the decay of the molecules is limited by the formation of the complex, and in fact, is suppressed with increasing $\gamma_j$.

We find that the dynamics of the particle decay is mainly determined by the inelastic part since the elastic collisions conserve the total particle number and only slightly affect the decay rate by redistributing the mode population (see details in \cite{supp}). Thus, in the following discussion, for simplicity we  set $g_{\text{re}} = 0$. In this case the corresponding rate equations simplify to 
\begin{eqnarray}
\frac{d \langle\hat{n}_{{\bm k}}\rangle}{dt}
&\approx&
-\sum_{{\bm k}'}\Gamma_{{\bm k}{\bm k}'}\,
\langle \hat{n}_{{\bm k}}\rangle\langle\hat{n}_{{\bm k}'}\rangle,
\label{eq:Nparticle}
\end{eqnarray}
with  $\Gamma_{{\bm k},{\bm k}'}=4g_{\text{im}}
|{\bm k}-{\bm k}'|^2/V$. The  complex population adiabatically follows the molecule population as $\langle \hat{n}^b_{{\bm k}}\rangle \approx
\sum_{{\bm k}'}\Gamma_{{\bm k}{\bm k}'}/(2\gamma)\,
\langle \hat{n}_{{\bm k}}\rangle\langle\hat{n}_{{\bm k}'}\rangle$.

We can easily generalise Eq.~(\ref{eq:Nparticle}) to any type of trapping potential $V(\bm r)$ by considering the corresponding single-particle eigenmodes.
Explicitly,
\begin{align}
\frac{d \langle\hat{n}_{{\bm n}}\rangle}{dt}
&
\approx-\sum_{{\bm n}'}\Gamma_{{\bm n}{\bm n}'}\langle\hat{n}_{{\bm n}}\rangle\langle\hat{n}_{{\bm n}'}\rangle,
\\
\frac{d N}{dt}&=
-\sum_{{\bm n}}\frac{d \langle\hat{n}_{{\bm n}}\rangle}{dt}
\equiv -\overline{\Gamma} N^2
\end{align}
where $\hat{n}_{{\bm n}}=\hat{c}^{\dagger}_{{\bm n}}\hat{c}_{{\bm n}}$ denotes the molecule population operator
in mode ${\bm n}$, $\Gamma_{{\bm n}{\bm n}'}$ is given by an integral over eigenmodes ${\bm n}$ and ${\bm n}'$ (see \cite{supp}), and we defined the time-dependent averaged particle decay rates as
$\overline{\Gamma}=\sum\Gamma_{{\bm n}{\bm n}'}\,\langle \hat{n}_{{\bm n}}\rangle
\langle\hat{n}_{{\bm n}'}\rangle/N^2$.

\textit{Comparison with Experiment }
We now apply this developed framework to the experimental conditions, assuming a $3$D harmonic trapping potential of the form $V({\bm r}) = \sum_{i=x,y,z} m\omega_{i}^2 r_i^2/2$
with $\omega_{i}$ the trapping frequency in the $i$-th direction.
We also assume $b_{\text{im}}=118 a_0$ ($a_0=5.29\times 10 ^{-11}$m) for $\text{KRb}$ molecules as calculated \cite{idziaszek2010universal} and experimentally verified in the classical temperature limit \cite{de2019degenerate}.
In addition to the total particle number $N(t)$, we study the density $n\equiv N/V$, the volume $V$, defined as $V=8\pi^{3/2}(\sigma_x\sigma_y\sigma_z)$ with $\sigma_i$ the standard deviation of the density profile in the $i$-th direction, the total energy $E$ and the energy density $\epsilon \equiv E/N$. 

\begin{figure}[tbh]
\centering
\includegraphics[width=0.45\textwidth]{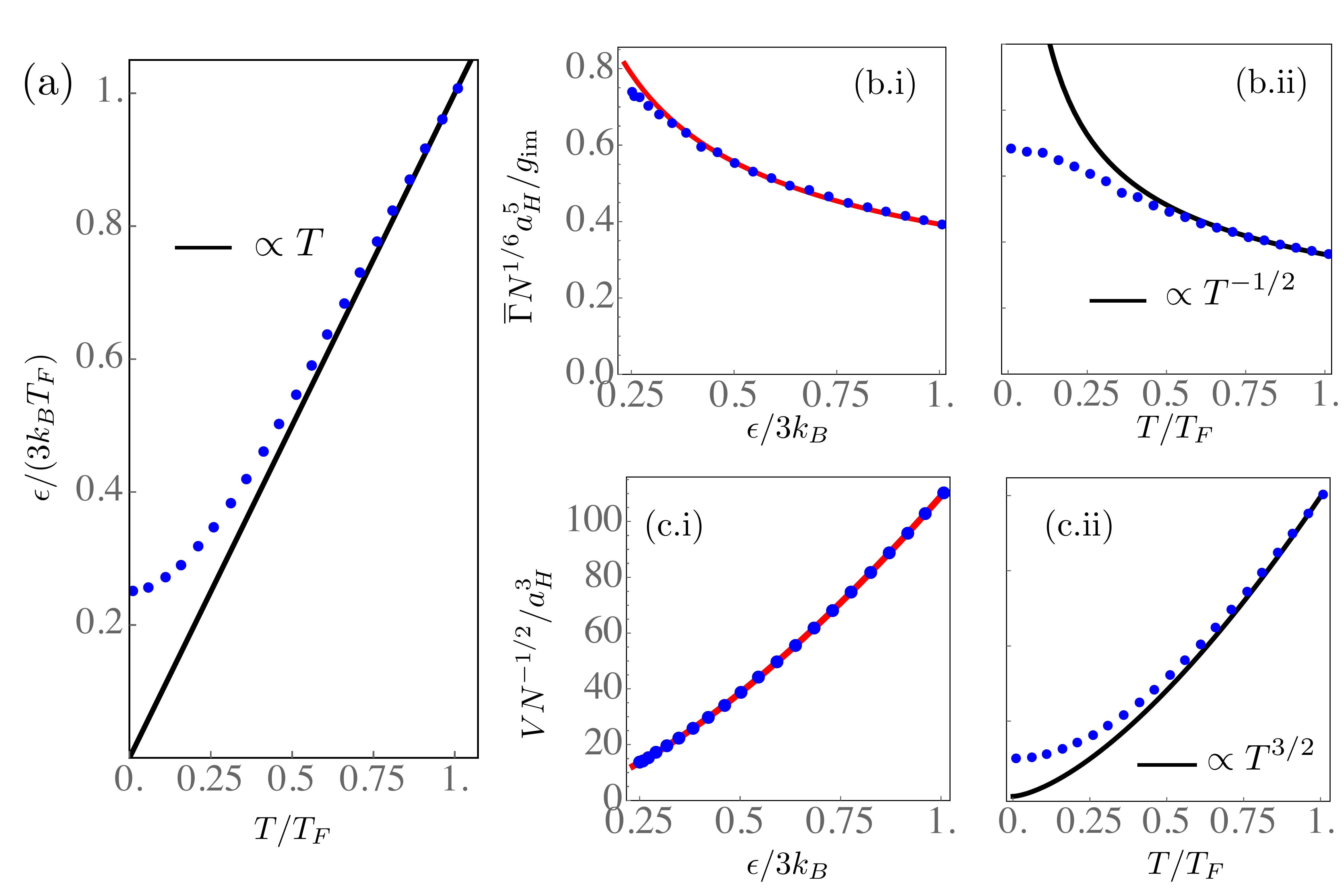}
\caption{
Thermodynamic scaling relations in a $3$D harmonic trap: (a) Average energy density, $\epsilon$, (b) $\overline{\Gamma}$ and (c) average volume, $V$.
Numerical results are shown as blue dots, and analytical scalings  as a function of energy density [red lines (bi,ci)] and temperature [black lines (bii,cii)]. Only the scaling in terms of $\epsilon$ remains valid in the quantum degenerate regime.}
\label{fig:Gamma_plot}
\end{figure}

To develop an analytical understanding, we explore the scaling relations of the averaged decay rates and the volume.
For an equilibrium system at temperature $T$,
the population $\langle\hat{n}_{\bf n}\rangle$ obeys the Fermi-Dirac distribution, from which the energy density $\epsilon$, $\overline{\Gamma}$ and $V$
can be obtained as a function of $T$.
As shown in Fig.~\ref{fig:Gamma_plot} (a), in the classical limit $T\gtrsim 0.5 T_F$,
the energy density $\epsilon$ in each direction is $k_BT$ in the harmonic trap in accordance with the equipartition theorem,
giving rise to $\epsilon = 3k_BT$;
in the quantum degenerate  limit $T\lesssim 0.5 T_F$, $\epsilon$ is higher than the one predicted by a classical scaling since the Fermi energy remains finite at zero temperature due to quantum statistics.
As demonstrated in Fig.~\ref{fig:Gamma_plot} (b),
the scaling relations, $\overline{\Gamma}\propto \epsilon^{-1/2}$ and $V \propto \epsilon^{3/2}$, written as a function of $\epsilon$, 
are universal over the whole temperature range, whereas only in the classical regime the replacement $\epsilon \to T$ is valid as shown in Fig.~\ref{fig:Gamma_plot} (b.ii) and (c.ii). 

During the non-equilibrium decay dynamics these simple relations derived in equilibrium are not necessarily  applicable. Notwithstanding, they are found to keep holding during the full dynamics as benchmarked by numerical simulations (see \cite{supp}). We attribute this partly to the fact that  in a harmonic trap the initial Fermi distribution remains approximately unchanged during the dynamics by the balancing between the local density and the p-wave decay rate: the low energy modes with a low p-wave decay rate concentrate at the trap center where the density is higher, while the high energy modes with faster decay rates,  concentrate at the edges where the density is lower, making the effective decay rate nearly uniform through the cloud.


In the experiment there is additional heating as particles are lost (see \cite{supp}) similar to the one   observed in prior experiments \cite{urvoy2019direct}. Here we  phenomenologically describe these heating processes by a background single particle heating rate $3k_Bh_{\text{bg}}$ acting as:
\begin{align}
\frac{dN}{dt} & = - \overline{\Gamma} N^2,\quad
\frac{d\epsilon}{dt} = 3k_B h_{\text{bg}},
\end{align}
where $V =V_0\big(\epsilon/\epsilon_0\big)^{3/2}$ and $\overline{\Gamma} = \overline{\Gamma}_0 \big(\epsilon/\epsilon_0\big)^{-1/2}$, and the subscript $0$ denotes the values at $t=0$.
The dynamics of the density $n(t)$ can be analytically obtained as
\begin{align}
n(t) &\approx\frac{n_0(1+3k_B h t/\epsilon_0)^{-3/2}}{1 + 2\overline{\Gamma}_0V_0n_0\sqrt{\epsilon_0}\big(\sqrt{\epsilon_0+3k_Bht}-\sqrt{\epsilon_0}\big)/(3k_Bh)}\, .
\label{eq:solution_theory}
\end{align}

%
%

From this expression   the density decay rate, which was the fitting parameter used to characterize the decay rate in Ref. \cite{de2019degenerate}, is predicted to be at short times $\beta_0\equiv\overline{\Gamma}_0V_0\propto\epsilon_0$, and thus  proportional to the energy density.   In the classical limit, it recovers the results of the Bethe-Wigner threshold law since $\epsilon=3k_BT$. In the quantum degenerate limit, the decay rate saturates to the Fermi energy $k_B T_F$ instead of decreasing to zero.


\begin{figure}[htb]
\centering
\includegraphics[width=0.35\textwidth]{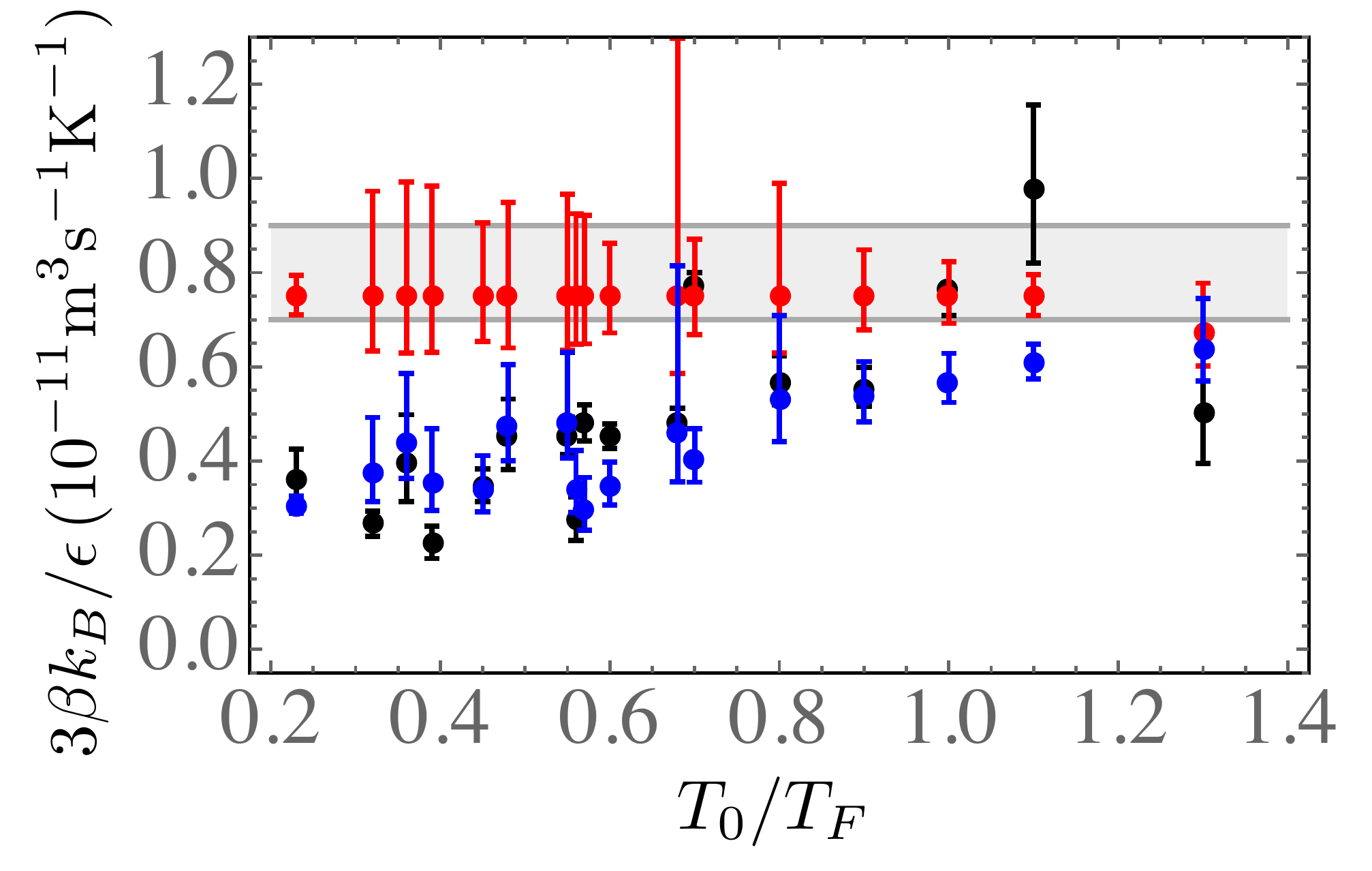}
\caption{Comparison of theory predictions considering pure p-wave molecule-molecule collisions without (red dots) or with (blue dots) additional complex-molecules collisions assuming $\alpha= 8\times 10^{-20}\text{m}^3$, and experimental measurements (black dots). Each dot corresponds to different experimental runs  with slightly different conditions (\cite{supp}).  The theory (experiment) $\beta_0/(\epsilon/3k_B)$ is obtained as the best fit of the theoretically derived (the experimentally measured) $n(t)$ to Eq.~(\ref{eq:solution_theory}).
The error bars include  uncertainties in the experimental measurements and the standard deviation from the  fitting procedure (see details in \cite{supp}).
In the classical temperature limit, both the theory and the experimental results are approximately constant, in agreement with the universal prediction \cite{idziaszek2010universal} indicated by the gray band accounting for $8\%$ errors in the scattering value $b_{\text{im}}^3$.
In the quantum degenerate limit, the model including the complex-molecule collisions can quantitatively reproduce the observed suppression.
}
\label{fig:HeatingAndB_plot}
\end{figure}
To directly  compare with the experimentally extracted rates, we numerically extract the  decay rate $\beta_0$  as the best fit of $n(t)$ to Eq.~(\ref{eq:solution_theory}) for the corresponding initial conditions (see \cite{supp} for detailed fitting procedures).
As shown in Fig.~\ref{fig:HeatingAndB_plot}, the theory resuls are flat throughout both the classical temperature and the quantum degenerate regime, while the experimental data shows a strong suppression in the latter. We note that both represent an enhancement compared to the classical expectation where the decay rate would vanish in the zero temperature limit.

\textit{Beyond two-body molecule loss}
Having established that pure two-body molecule decay is insufficient to explain the experimentally observed suppression, we consider more complicated  interaction processes in our model. %
A change of the effective loss rate in our framework requires either the coherent coupling of molecules to complexes or the incoherent decay of the complexes themselves to be modified. Modifications to the former may arise from higher-order elastic interactions between molecules    during  complex formation, while modifications to the later could arise from inelastic complex-molecule decay channels due to for example   light-assisted collisions. In fact, recent measurements of complex decay rates in KRb  \cite{liu2020steering} as well as other non-reactive molecules \cite{Cornish} observed significant enhancement of the complex decay rate via photo-excitation processes.%
In the following we explore such type of (in)elastic collisions between molecules and the complex, as illustrated in Fig.~\ref{fig:schematic_plot}(c).  
To connect to Eq.~(\ref{eq:TwoChannel}), for simplicity we start again by considering a homogeneous gas and model these processes by adding the following terms in the master equation
\begin{align}
\hat{H}_{\rm {int}}' &=\sum_{j, {\bm k}, {\bm k}',{\bm k}''}\frac{\hbar\alpha_{g} }{2 V}G^j_{{\bm k},{\bm k}'}
 \big(
\hat{b}^{\dagger}_{j,{\bm k}'+{\bm k}''}\hat{c}_{{\bm k}}^{\dagger}\hat{c}_{{\bm k}}\hat{c}_{{\bm k}'}\hat{c}_{{\bm k}''}
+\text{h.c.}
\big), \notag\\
\mathcal{L}'(\hat{\rho})&=\alpha_{\gamma}\sum_{j,{\bm k},{\bm k}'}\gamma_j
\mathcal{L}[\hat{b}_{j,{\bm k}}\hat{c}_{{\bm k}'}]\,\hat{\rho}/V,
\label{eq:three-body}
\end{align} where the parameters $\alpha_{g}G^j_{{\bm k},{\bm k}'}/V$ and $\alpha_{\gamma}\gamma_j/V$, which have the unit of $\text{s}^{-1}$, parametrizes the rates of three body elastic collisions and the molecule-complex decay respectively.  
After adiabatically eliminating the complex, these terms result in a modification of the two-body decay rate and an additional molecular three-body  decay term  \cite{supp}
\begin{equation}
\begin{split}
\frac{d \langle\hat{n}_{{\bm k}}\rangle}{dt}
\approx 
&-\sum_{{\bm k}'}\frac{\big(1+2\alpha_{g}n\big)}{\big(1+\alpha_{\gamma}n\big)}\Gamma_{{\bm k},{\bm k}'}
\langle\hat{n}_{{\bm k}}\rangle\langle\hat{n}_{{\bm k}'}\rangle\\
& -\sum_{j,{\bm k}',{\bm k}''}\frac{\alpha_{\gamma}}{2V}
\Gamma_{{\bm k}',{\bm k}''}
\langle\hat{n}_{{\bm k}}\rangle\langle\hat{n}_{{\bm k}'}\rangle\langle\hat{n}_{{\bm k}''}\rangle,
\end{split}
\label{eq:nparticleNew}
\end{equation}
where $n=\sum_{{\bm k}''}\langle \hat{n}_{{\bm k}''}\rangle/V=N/V$ is the density of the molecular gas and we assumed $E_j\ll \gamma_j$.
Consequently, the total number of molecules follows 
\begin{eqnarray}
\frac{d N}{dt}
&\approx&-\sum_{{\bm k},{\bm k}'}\Gamma_{{\bm k},{\bm k}'}^{P}
\langle\hat{n}_{{\bm k}}\rangle\langle\hat{n}_{{\bm k}'}\rangle,
\label{eq:NparticleNew}
\end{eqnarray}
where the modified decay rate becomes $\Gamma_{{\bm k},{\bm k}'}^{P}\equiv\Gamma_{{\bm k},{\bm k}'}\big(1+(2\alpha_{g} -\alpha_{\gamma}/2)\, n\big)$, with an effective inelastic scattering parameter $g_{\text{im}}^{P}\equiv g_{\text{im}}\big(1-\alpha n \big)$ and $\alpha = \alpha_{\gamma}/2 - 2\alpha_{g}$.
Thus, counter-intuitively the additional loss due to complex-molecule collisions results in an effective suppression of the two-body loss due to the quantum Zeno effect which suppresses the population of the complex for larger loss rates.

For a  gas trapped in a $3$D harmonic potential this density dependent scattering strength $g_{\text{im}}^{P}$ generates an effective loss suppression in the quantum degenerate regime if $\alpha>0$ as the gas becomes denser with decreasing temperature. In a system with a fixed particle number, where the change in density is directly correlated with the average  volume, this suppression of the decay rate is tied to the temperature dependence of the average volume, see Fig.~\ref{fig:Gamma_plot}(c), reflecting the underlying Fermi statistics.

In Fig.~\ref{fig:HeatingAndB_plot} we demonstrate that this effective model can reproduce the experimentally observed suppression when choosing $\alpha=8\times 10^{-20}\text{m}^3$. However, we note that this corresponds to an inelastic collision rate $\alpha_{\gamma} \gamma$ between molecules and the complex which exceeds the unitary limit. In contrast, the elastic term $\alpha_g g$ is in principle feasible, but requires a coherent three body process, rather than the conventionally expected pure loss in a three-body collision \cite{Esry1999,Esry2001}. A full explanation of  the underlying many-body framework responsible for the emergence of this terms, either  from quasi-particle dressing and in medium interactions,   or direct  multi-body  or light assisted collisions is still pending.

\textit{Conclusions and outlook} 
We have developed a theoretical framework that accounts for the formation of an intermediate molecular complex to study the reactive  dynamics of a quantum degenerate gas of polar molecules. 
The first part of this work considering pure p-wave collision of the molecules establishes a decay rate proportional to the energy density of the gas, extending the classical Wigner threshold law to the quantum degenerate regime, and predicts a flat behaviour at low temperature enhanced compared to the linearly in T vanishing classical prediction. 
However, as two-body molecule decay processes  mediated  by  the formation of complexes alone does not reproduce the experimentally observed behaviour in the quantum degenerate regime, we considered  beyond two-body molecule collisions. By including elastic or inelastic  higher order  complex-molecule interactions  we are  able  to reproduce the experimental observations. Nevertheless, it seems unlikely that  the actual origin  of these terms are direct complex-molecule  collisions. Instead they may emerge from many-body effects in the presence of trapping light. We hope that our conclusions can stimulate further theory work understanding the microscopic origin of these effects and experimental work that can  directly validate or refute  our predictions.



\begin{acknowledgements}
We acknowledge helpful discussions with Jun Ye and his JILA KRb group,  Joseph Thywissen, Paul Julienne, John Bohn and Qi Zhou during the preparation of this manuscript. This work is supported by the ARO single investigator award W911NF-19-1-0210, the DARPA DRINQs program and the  JILA-PFC PHY-1734006 grants, by NIST, and by NSF grant PHY--1912350.
\end{acknowledgements}


\widetext

\section{Derivations of the rate equations}
\label{App:rate_equations}

Here we first derive   the equations of motions described by Eq.~(1-4) in the main text.  
We define$\hat{A}_{{\bm k},{\bm k}'}=\hat{a}_{{\bm k}}\,\hat{a}_{{\bm k}'}$ and
$\hat{C}_{{\bm k},{\bm k}'}=\hat{a}^{\dagger}_{{\bm k}}\,\hat{a}_{{\bm k}'}$)

\begin{eqnarray}
\frac{d \,\langle\hat{b}_{{\bm k},j}^{\dagger}\hat{b}_{{\bm k}',j'}\rangle}{dt} &=&
 i \big(-E_{j} +E_{j'}+ \frac{|{\bm k}'|^2-|{\bm k}|^2}{4m}\big) 
\langle\hat{b}_{{\bm k},j}^{\dagger}\hat{b}_{{\bm k}',j'}\rangle
-(\gamma_{j}+\gamma_{j'})\langle\hat{b}_{{\bm k},j}^{\dagger}\hat{b}_{{\bm k}',j'}\rangle
\nonumber\\
&&+\frac{i}{\sqrt{V}}\sum_{{\bm k}''}\Big(
g_{j'} |2{\bm k}''-{\bm k}'| \,\langle\hat{b}_{{\bm k},j}^{\dagger}\hat{A}_{{\bm k}'',{\bm k}'-{\bm k}''}\rangle
 - g_j |2{\bm k}''-{\bm k}| \,\langle\hat{A}_{{\bm k}'',{\bm k}-{\bm k}''}^{\dagger}\hat{b}_{{\bm k}',j'}\rangle
 \Big)
\label{eq:bbField}
\end{eqnarray}

\begin{eqnarray}
\frac{d \langle\hat{A}_{{\bm k},{\bm k}'}\rangle}{dt} &=&
i \frac{|{\bm k}|^2+|{\bm k}'|^2}{2m} \langle\hat{A}_{{\bm k},{\bm k}'}\rangle
+i\frac{2}{\sqrt{V}}\sum_{j}g_j|{\bm k}-{\bm k}'|\,\langle\hat{b}_{{\bm k}+{\bm k}',j}\rangle
\nonumber\\
&&+i\frac{2}{\sqrt{V}}\sum_{{\bm k}'',j}g_j
\Big(
|{\bm k}''-{\bm k}|\,\langle\hat{C}_{{\bm k}'',{\bm k}'} \, \hat{b}_{{\bm k}''+{\bm k},j}\rangle
+|{\bm k}'-{\bm k}''|\,\langle\hat{C}_{{\bm k}'',{\bm k}} \, \hat{b}_{{\bm k}''+{\bm k}',j}\rangle
\Big)
\label{eq:aField}
\end{eqnarray}

\begin{eqnarray}
\frac{d \,\langle\hat{b}_{{\bm k}'',j}^{\dagger}\hat{A}_{{\bm k},{\bm k}'}\rangle}{dt} &=&
i \Big(\frac{2|{\bm k}|^2+2|{\bm k}'|^2-|{\bm k}''|^2}{4m} -E_{j}  \Big)
\langle\hat{b}_{{\bm k}'',j}^{\dagger}\hat{A}_{{\bm k},{\bm k}'}\rangle
-\gamma_{j}\langle\hat{b}_{{\bm k}'',j}^{\dagger}\hat{A}_{{\bm k},{\bm k}'}\rangle
\nonumber\\
&&+i\frac{2}{\sqrt{V}}\sum_{j'}g_{j'}|{\bm k}-{\bm k}'|\,\langle\hat{b}_{{\bm k}'',j}^{\dagger}\hat{b}_{{\bm k}+{\bm k}',j'}\rangle
-i\frac{2}{\sqrt{V}}\sum_{{\bm k}'''}g_j|2{\bm k}'''-{\bm k}''|
\langle\hat{A}_{{\bm k}''',{\bm k}''-{\bm k}'''}^{\dagger}\hat{A}_{{\bm k},{\bm k}'}\rangle
\nonumber\\
&&+i\frac{2}{\sqrt{V}}\sum_{{\bm k}''',j'}g_{j'}
\Big(
|{\bm k}'''-{\bm k}|\,\langle\hat{b}_{{\bm k}'',j}^{\dagger}\hat{C}_{{\bm k}''',{\bm k}'}\hat{b}_{{\bm k}'''+{\bm k},j'}\rangle
+|{\bm k}'-{\bm k}'''|\,\langle\hat{b}_{{\bm k}'',j}^{\dagger}\hat{C}_{{\bm k}''',{\bm k}}\hat{b}_{{\bm k}'''+{\bm k}',j'}\rangle
\Big)
\label{eq:baField}
\end{eqnarray}

\begin{eqnarray}
\frac{d \langle\hat{C}_{{\bm k},{\bm k}'}\rangle}{dt}
&=&
i \frac{|{\bm k}'|^2-|{\bm k}|^2}{2m} \langle\hat{C}_{{\bm k},{\bm k}'}\rangle
+
i\frac{2}{\sqrt{V}}\sum_{j,{\bm k}''}g_j\Big(
|{\bm k}-{\bm k}''|\,\langle\hat{b}^{\dagger}_{{\bm k}''+{\bm k},j} \, \hat{A}_{{\bm k}'',{\bm k}'}\rangle
+|{\bm k}''-{\bm k}'|\,\langle\hat{A}^{\dagger}_{{\bm k}'',{\bm k}}\,\hat{b}_{{\bm k}''+{\bm k}',j}\rangle
\Big)
\nonumber\\
\label{eq:aNField}
\end{eqnarray}

\begin{eqnarray}
\frac{d \,\langle\hat{C}_{{\bm k},{\bm k}'}\hat{b}_{{\bm k}'',j}\rangle}{dt} &=&
i \Big(\frac{2|{\bm k}'|^2-2|{\bm k}|^2+|{\bm k}''|^2}{4m} +E_{j}+i\gamma_j  \Big)
\langle\hat{b}_{j,{\bm k}''}\hat{C}_{{\bm k},{\bm k}'}\rangle
+i\frac{2}{\sqrt{V}}\sum_{{\bm k}'''}g_j|2{\bm k}'''-{\bm k}''|
\langle\hat{C}_{{\bm k},{\bm k}'}\hat{A}_{{\bm k}''',{\bm k}''-{\bm k}'''}\rangle
\nonumber\\
&&
+
i\frac{2}{\sqrt{V}}\sum_{j',{\bm k}'''}g_{j'}\Big(
|{\bm k}-{\bm k}'''|\,\langle\hat{b}^{\dagger}_{{\bm k}'''+{\bm k},j'} \, \hat{A}_{{\bm k}''',{\bm k}'}\,\hat{b}_{{\bm k}'',j}\rangle
+|{\bm k}'''-{\bm k}'|\,\langle\hat{A}^{\dagger}_{{\bm k}''',{\bm k}} \, \hat{b}_{{\bm k}'''+{\bm k}',j'}\,\hat{b}_{{\bm k}'',j}\rangle
\Big)
\label{eq:bcField}
\end{eqnarray}

Assuming the observables can be factorized as
\begin{eqnarray}
\langle\hat{A}_{{\bm k}_1,{\bm k}_2}\hat{A}_{{\bm k}_3,{\bm k}_4}^{\dagger}\rangle
&=&
\langle\hat{A}_{{\bm k}_1,{\bm k}_2}\rangle\langle\hat{A}_{{\bm k}_3,{\bm k}_4}^{\dagger}\rangle
-\big(\langle\hat{C}_{{\bm k}_4,{\bm k}_1}^{\dagger}\rangle -\delta_{{\bm k}_1,{\bm k}_4} \big)
\big(\langle\hat{C}_{{\bm k}_3,{\bm k}_2}\rangle -\delta_{{\bm k}_2,{\bm k}_3} \big)
\nonumber\\
&&+\big(\langle\hat{C}_{{\bm k}_3,{\bm k}_1}^{\dagger}\rangle -\delta_{{\bm k}_1,{\bm k}_3} \big)
\big(\langle\hat{C}_{{\bm k}_4,{\bm k}_2}\rangle -\delta_{{\bm k}_2,{\bm k}_4} \big)
\\
\langle\hat{C}_{{\bm k}_1,{\bm k}_2}\hat{A}_{{\bm k}_3,{\bm k}_4}^{\dagger}\rangle
&=&
\langle\hat{C}_{{\bm k}_1,{\bm k}_2}\rangle\langle\hat{A}_{{\bm k}_3,{\bm k}_4}^{\dagger}\rangle
-\big(\langle\hat{C}_{{\bm k}_4,{\bm k}_1}^{\dagger}\rangle -\delta_{{\bm k}_1,{\bm k}_4} \big)
\big(\langle\hat{C}_{{\bm k}_3,{\bm k}_2}\rangle -\delta_{{\bm k}_2,{\bm k}_3} \big)
\nonumber\\
&&+\big(\langle\hat{C}_{{\bm k}_3,{\bm k}_1}^{\dagger}\rangle -\delta_{{\bm k}_1,{\bm k}_3} \big)
\big(\langle\hat{C}_{{\bm k}_4,{\bm k}_2}\rangle -\delta_{{\bm k}_2,{\bm k}_4} \big),
\end{eqnarray}
then the equations of motion above become a closed set of equations and the dynamics of the observables can be evaluated. Our numerical simulations confirm  the coherence terms $\langle\hat{A}_{{\bm k}_1,{\bm k}_2}\rangle$,  $\langle\hat{C}_{{\bm k}_1,{\bm k}_2}\rangle$ (${\bm k}_1\neq {\bm k}_2$) and $\langle\hat{b}_{{\bm k}_3,j}^{\dagger}\hat{A}_{{\bm k}_1,{\bm k}_2}\rangle$ (${\bm k}_3\neq {\bm k}_1+{\bm k}_2$), which are initially zero, remain zero, and therefore can be neglected. Then the relevant observables are  $\langle\hat{C}_{{\bm k}_1,{\bm k}_1}\rangle$, $\langle\hat{b}_{{\bm k}_j}^{\dagger}\hat{b}_{{\bm k}_{j'}}\rangle$ 
and $\langle\hat{b}_{{\bm k}_1+{\bm k}_2,j}^{\dagger}\hat{A}_{{\bm k}_1,{\bm k}_2}\rangle$.

In  Ref. \cite{liu2020steering}, the lifetime of the KRb complex was measured to be  $\gtrsim 250$ns,
indicating $\gamma\gtrsim 2\pi \times 4$MHz.
Since the experimentally relevant  energy scales  are set by  the Fermi energy ($\sim$ kHz),  which is  much smaller than the complex decay rate,   we can  adiabatically eliminate the complex, and set to zero both the left hand side of Eq.~(\ref{eq:baField}) and the term  
$\langle\hat{b}_{j,{\bm k}+{\bm k}'}^{\dagger}\hat{b}_{j,{\bm k}+{\bm k}'}\rangle$. 
Then the correlation terms can be approximated as
\begin{eqnarray}
\langle\hat{b}_{j,{\bm k}+{\bm k}'}^{\dagger}\hat{A}_{{\bm k},{\bm k}'}\rangle
&\approx&
-i\frac{4g_j}{\sqrt{V}}|{\bm k}-{\bm k}'|
\langle\hat{n}_{{\bm k}}\hat{n}_{{\bm k}'}\rangle/(\gamma-iE_{j})
\approx
-i\frac{4g_j}{\sqrt{V}}|{\bm k}-{\bm k}'|
\langle\hat{n}_{{\bm k}}\rangle\langle\hat{n}_{{\bm k}'}\rangle/(\gamma-iE_{j}),
\end{eqnarray}
where $\hat{n}_{\bm k}=\hat{c}_{\bm k}^{\dagger}\hat{c}_{\bm k}$ and the second approximation is taken since the coherence term is zero. There we have also ignored single particle kinetic energy terms since they are in the order of $\sim \text{kHz} \ll \gamma_j\sim \text{MHz}$. 
By substituting the correlations into Eq.~(\ref{eq:aNField}), the dynamics for the molecular population becomes
\begin{eqnarray}
\frac{d \langle\hat{n}_{{\bm k}}\rangle}{dt}
&=&
-\sum_{{\bm k}'}\Gamma_{{\bm k},{\bm k}'}\langle \hat{n}_{{\bm k}}\rangle\langle \hat{n}_{{\bm k}'}\rangle,
\quad
\Gamma_{{\bm k},{\bm k}'}=\sum_j\frac{16g_j^2}{V}\frac{\gamma_j}{\gamma_j^2+E_j^2}|{\bm k}-{\bm k}'|^2.
\label{eq:rate}
\end{eqnarray}

Note that Eq.~(\ref{eq:rate}) recovers the standard rate equations that describe direct chemical reactions, if we identify  $g_{\text{im}}\equiv 3 \pi \hbar b_{\text{im}}^3/m= \sum_j 4g_j^2 \gamma_j/(\gamma_j^2+E_j^2)$ and $g_{\text{re}}\equiv  3 \pi \hbar b_{\text{re}}^3/m= \sum_j 4g_j^2 E_j/(\gamma_j^2+E_j^2)$, with $ b_{\text{re}}^3$ and $b_{\text{im}}^3$ the real and imaginary parts of the scattering volume.

As mentioned in the main text, Eq.~(\ref{eq:rate}) can be generalized to account for any type of trapping potentials $V(\bm r)$ by replacing $\langle\hat{n}_{{\bm k}}\rangle$ by the population $\langle\hat{n}_{{\bm n}}\rangle$ as given by
\begin{eqnarray}
\frac{d \langle\hat{n}_{{\bm n}}\rangle}{dt}
&\approx-\sum_{{\bm n}'}\Gamma_{{\bm n}{\bm n}'}\langle\hat{n}_{{\bm n}}\rangle\langle\hat{n}_{{\bm n}'}\rangle,
\label{eq:rateHarmonic}
\end{eqnarray}
where $\Gamma_{{\bm n}{\bm n}'}\equiv4\Gamma_{{\bm n}{\bm n}'{\bm n}{\bm n}'}$ with 

\begin{eqnarray}
\Gamma_{{\bm n}{\bm n}'{\bm n}''{\bm n}'''}=\frac{3\pi \hbar b_{\text{im}}^3 }{m}\Big(\int d{\bm r}^{\,3}
\big[\big({\bm \nabla}\phi^{*}_{{\bm n}}(\bm r)\big)\phi^{*}_{{\bm n}'}(\bm r)
-\phi^{*}_{{\bm n}}(\bm r)\big({\bm \nabla}\phi^{*}_{{\bm n}'}(\bm r)\big)\big]
\cdot\big[\big({\bm \nabla}\phi_{{\bm n}''}(\bm r)\big)\phi_{{\bm n}'''}(\bm r)
-\phi_{{\bm n}''}(\bm r)\big({\bm \nabla}\phi_{{\bm n}'''}(\bm r)\big)\big]\Big),
\nonumber\\
\label{eq:GammaIndex}
\end{eqnarray}
where $\phi_{{\bm n}}(\bm r)$ is the eigenfunction of the eigenmode ${\bm n}$ of the single particle Hamiltonian.

\section{Effects of the elastic scattering}
\label{App:elastic}
\begin{figure}[hbt!]
\centering
\includegraphics[width=0.85\textwidth]{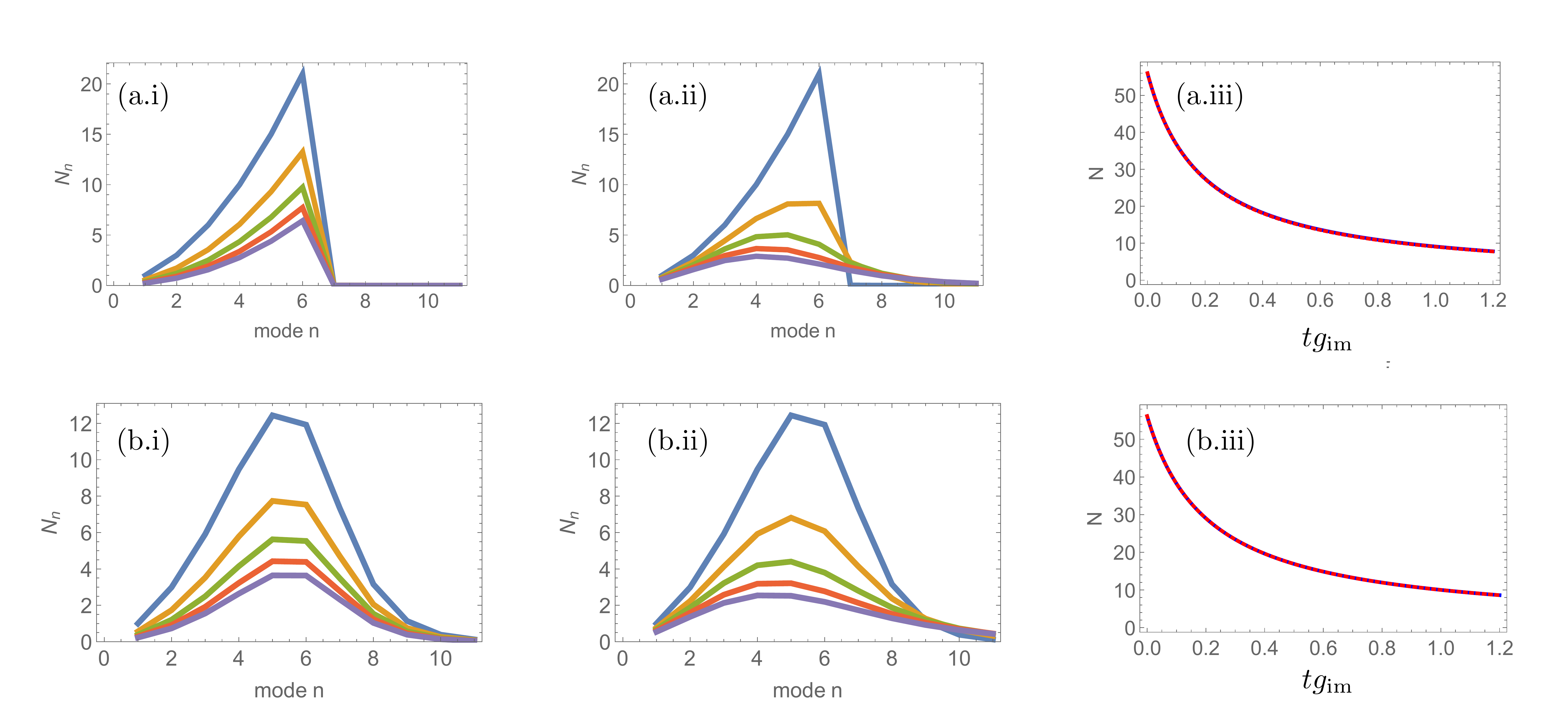}
\caption{Population dynamics for different   elastic scattering volumes  and  temperatures: (a) $T=0$, (b) $T=0.3 T_F$.
Panel (i) and panel (ii) show the particle mode  distribution for $N(t=0)=56$ particles  ($n=n_x+n_y+n_z$) for  $g_{\text{re}}=0$ and $g_{\text{re}}=-g_{\text{im}}$ respectively.
The different colors represent the distribution at different times $t$: blue: $t=0$, yellow: $t=0.12 g_{\text{im}}$, 
green: $t=0.24 g_{\text{im}}$, red: $t=0.32 g_{\text{im}}$ and purple: $t=0.48 g_{\text{im}}$.
Panel (iii) plot the  dynamics of the particle number $N(t)$ as a function of time (red: $g_{\text{re}}=0$, blue dashed: $g_{\text{re}}=-g_{\text{im}}$).The comparison shows that the elastic collisions only slightly affect the decay rate by redistributing the density profile and do not affect the decay dynamics.
To account for the fact that our simulations can not be done for large systems, we capture  the effect  of the elastic interactions expected for   the real particle number used in the experiment $N_{\text{exp}}=10^4$, by  rescaling  both $g_{\text{re}}$ and $g_{\text{im}}$ by a factor of $(N_{\text{exp}}/56)^{-1/6}\approx 0.42$, given 
the known   scaling of $\Gamma\propto N^{-1/6}$.
}
\label{fig:elastic_collision}
\end{figure}

As discussed in the last section, both  elastic  and inelastic interactions are    present. 
According to the multichannel quantum defect theory (MQDT) \cite{idziaszek2010universal},
the elastic and inelastic  scattering volumes in KRb  have exactly the same amplitude but with opposite sign. 
However, thermalization  effect of the elastic collision cannot be captured by  a second order  cumulant expansion such as the one used to derive  Eq.~(\ref{eq:aNField})
Instead here we use  kinetic theory to incorporate thermalization processes induced by elastic collisions  \cite{holland1997bose}  and demonstrate that for the case of KRb they play a minimal role in the  loss dynamics. In the context of kinetic theory the rate equations  read:

\begin{eqnarray}
\frac{d\langle \hat{n}_{{\bm n}}\rangle}{dt}&=&
-\sum_{{\bm n}'}\Gamma_{{\bm n}{\bm n}'}\langle\hat{n}_{{\bm n}}\rangle\langle\hat{n}_{{\bm n}'}\rangle
+\sum_{{\bm n}'{\bm n}''{\bm n}'''}W_{{\bm n}{\bm n}'{\bm n}''{\bm n}'''}\,\Big(\langle\hat{n}_{{\bm n}''}\rangle\,\langle\hat{n}_{{\bm n}'''}\rangle\,(1-\langle\hat{n}_{{\bm n}}\rangle)(1-\langle\hat{n}_{{\bm n}'}\rangle)-\langle\hat{n}_{{\bm n}}\rangle\,\langle\hat{n}_{{\bm n}'}\rangle\,(1-\langle\hat{n}_{{\bm n}''}\rangle)(1-\langle\hat{n}_{{\bm n}'''}\rangle)\Big),
\nonumber\\
\label{eq:elastic_rate}
\end{eqnarray}
where $W_{{\bm n}{\bm n}'{\bm n}''{\bm n}'''}=2\pi/\omega|g_{\text{re}}/g_{\text{im}}|^2 |\Gamma_{{\bm n}{\bm n}'{\bm n}''{\bm n}'''}|^2\delta_{E_{{\bm n}}+E_{{\bm n}'},E_{{\bm n}''}+E_{{\bm n}'''}}$ and $E_{{\bm n}}$ is the single particle energy of  mode ${\bm n}$.
Note that even though the elastic collisions are responsible  for  thermalization, the loss dynamics 
 is mainly determined by the inelastic part since the elastic collisions conserve the total particle number and only slightly affect the decay rate by redistributing the mode population, as shown in Fig.~\ref{fig:elastic_collision}.
 
\section{Evaporative heating}
\label{App:evaporative_heating}

\begin{figure}[hbt!]
\centering
\includegraphics[width=0.25\textwidth]{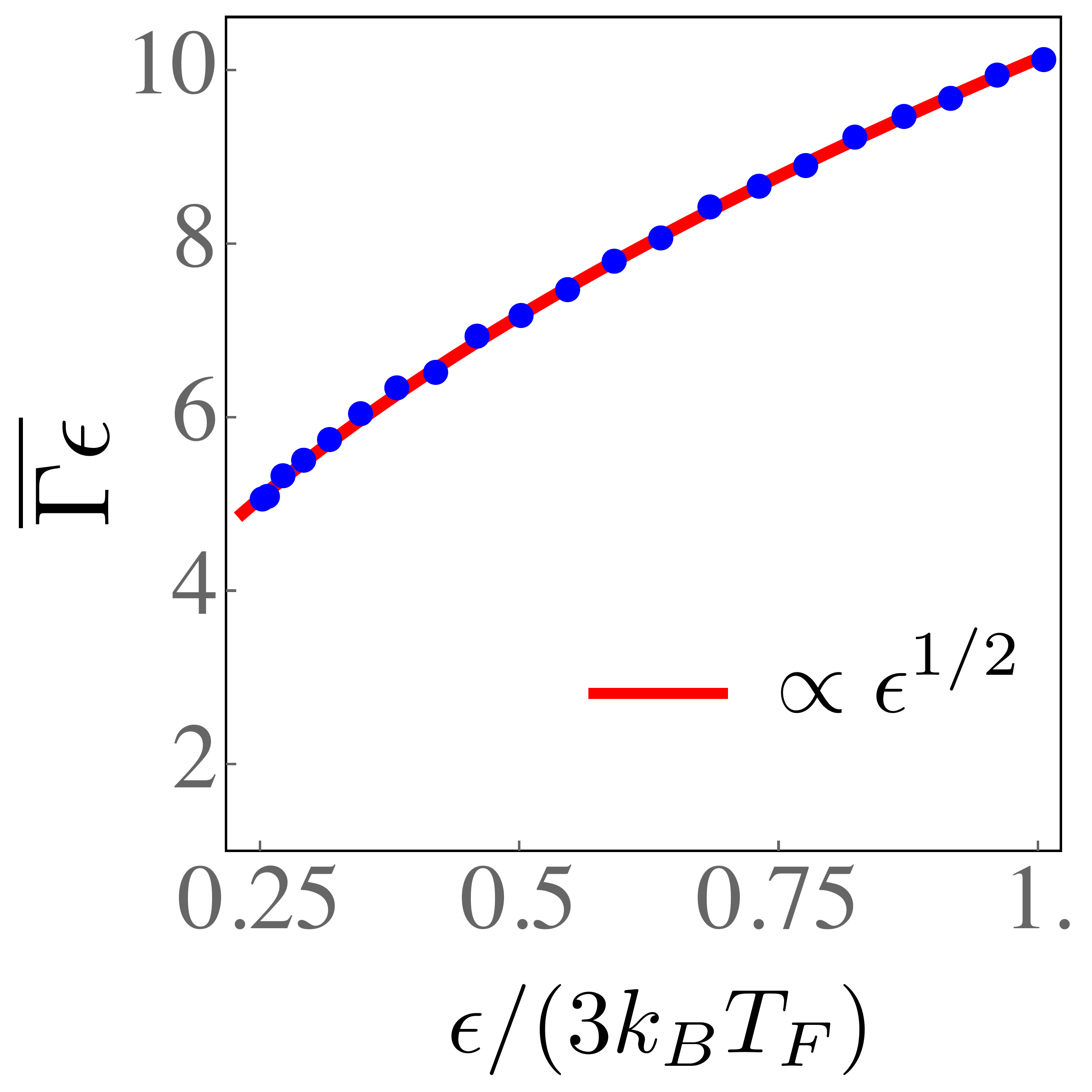}
\caption{ The scaling relation $\overline{\Gamma\epsilon}\propto \epsilon^{1/2}$ is valid over a wide range of $\epsilon$ that covers both  the classical limit and the quantum  degenerate regime.}
\label{fig:GammaE}
\end{figure}

During the decay process, the evolution of the total energy of the system is given by:
\begin{align}
\frac{d E(t)}{dt}&=
-\sum_{{\bm n}_i, {\bm n}_j} E_{{\bm n}_i}\Gamma_{{\bm n}_i{\bm n}_j}\langle\hat{n}_{i}\rangle_t\,\langle\hat{n}_{j}\rangle_t
\equiv-\overline{\Gamma\epsilon}(t)N(t)^2,
\label{eq:dynamicsE}
\end{align}
where the time-dependent averaged particle decay rate is defined as
$\overline{\Gamma\epsilon}(t)=\sum_{{\bm n}_i, {\bm n}_i}\Gamma_{{\bm n}_i{\bm n}_j} E_{{\bm n}_i}\,\langle\hat{n}_{i}\rangle_t\langle\hat{n}_{j}\rangle_t/N(t)^2$.
This equation  together with the dynamics of $N(t)$, can be used to  solve for 
the dynamics of the energy density which evolves  as 
\begin{align}
d\epsilon(t)/dt &= N(\overline{\Gamma}(t)\epsilon(t)-\overline{\Gamma\epsilon}(t))
\equiv\alpha_0 N\overline{\Gamma}\epsilon,
\end{align}
where $\alpha_0\equiv (\overline{\Gamma}\epsilon-\overline{\Gamma\epsilon})/\overline{\Gamma}\epsilon$
denotes the evaporative cooling(heating) rate with negative(positive) value.
For the $3$D harmonic confinement under consideration,
 the particles with lower energy decay faster according to the scaling $\Gamma\propto \epsilon^{-1/2}$. Therefore 
the energy density increases as particles get lost and the system is evaporatively heated up.
Using the numerical results in Fig.~\ref{fig:GammaE},
$\alpha_0$ is found to be  a constant $\alpha_0=0.07$ for all regimes down to $T=0.2T_{F}$ 
which is close to the result $\alpha_0 = 1/12$ in the classical regime predicted in \cite{ni2010dipolar} using a kinetic theory formalism.

\section{Simplified analytical equations}
\label{App: comparison to the equilibrium case}

\begin{figure}[hbt!]
\centering
\includegraphics[width=0.35\textwidth]{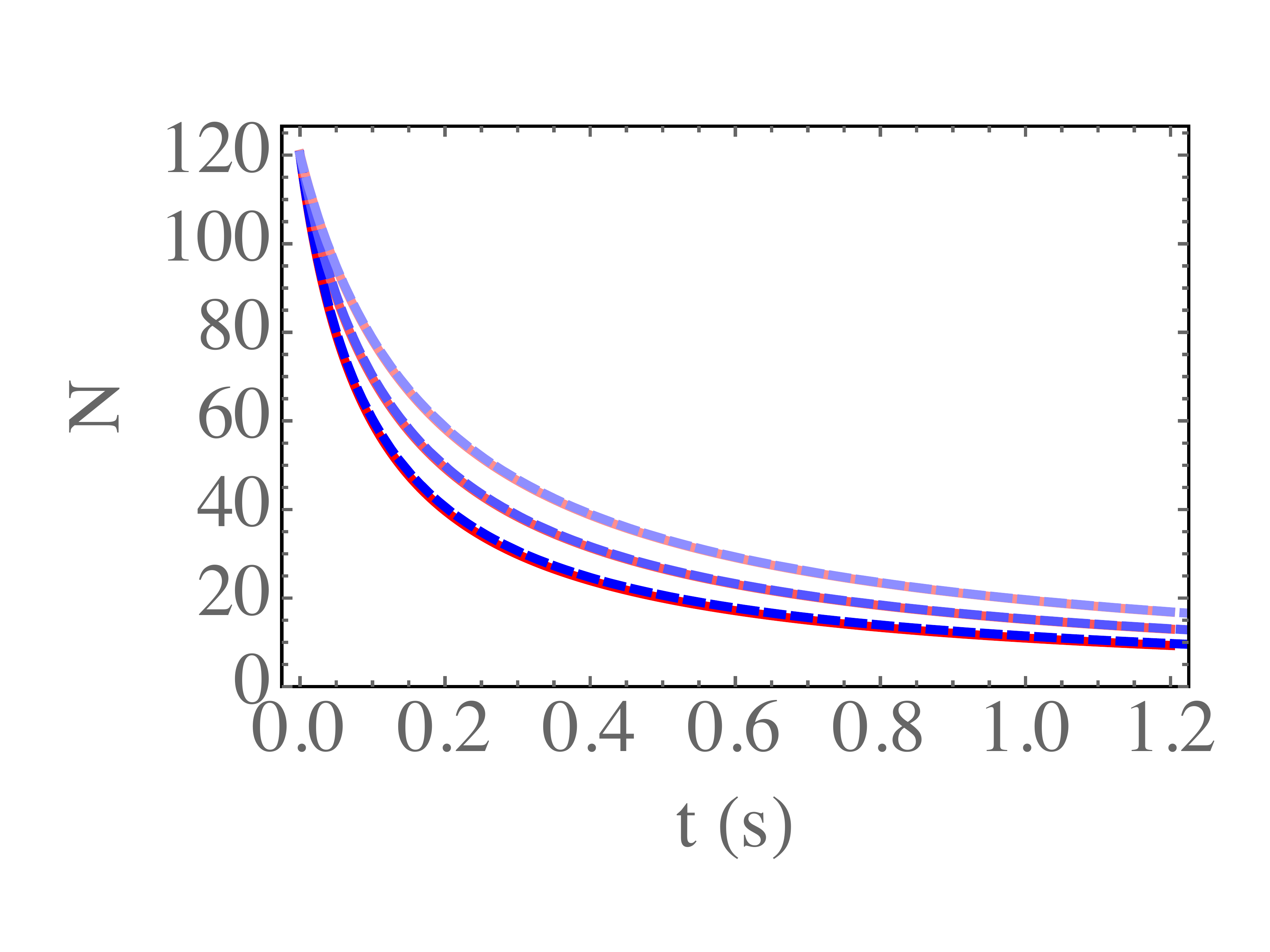}
\caption{ Comparisons  between the analytical results (red curves) and the numerical results (blue dashed lines) for the population dynamics at different initial equilibrium temperatures $T=0$, $T=0.5 T_F$ and $T=1.0 T_F$ from bottom to top for $N=120$ particles.
We find that the analytical results can well capture the numerically obtained dynamics over a wide range  of  temperatures. 
}
\label{fig:benchmark}
\end{figure}

The experiment measured the decay dynamics of an ensemble of $N\sim  10^5$ particles, for which a quantitative theoretical comparison is numerically hard, even at the mean-field level.  
To overcome this numerical complexity as well as getting more insight into the decay,
we assume that the decay dynamics is governed by simple analytical equations  which are  valid when the system is in equilibrium. 
Surprisingly, by performing comparisons with  numerical calculations we find that these relations describe well the  decay dynamics as shown in Fig.~\ref{fig:benchmark}

\section{Incorporating complex-molecule collisions in the rate equations}
\label{App:rate_equations2}

Taking into account the complex-molecule collisions, the equations of motions for the relevant observables become
\begin{eqnarray} 
\frac{d \langle\hat{n}_{{\bm k}}\rangle}{dt}
&=&
-\Big(\sum_{j,{\bm k}'}2\alpha_{\gamma}\gamma_j/V\langle\hat{n}^{b}_{j,{\bm k}'}\rangle\Big)
\langle\hat{n}_{{\bm k}}\rangle
+\sum_{j,{\bm k}'}\frac{4g_j}{\sqrt{V}}|{\bm k}-{\bm k}'|\,
\text{Im}[\langle\hat{b}^{\dagger}_{j,{\bm k}+{\bm k}'} \, \hat{A}_{{\bm k},{\bm k}'}\rangle]\nonumber\\
&&+\alpha_g\sum_{{\bm k}''}\text{Im}[\langle\hat{b}^{\dagger}_{j,{\bm k}+{\bm k}'} \, \hat{A}_{{\bm k},{\bm k}'}\hat{n}_{{\bm k}''}\rangle]/V
\label{eq:ccFieldNew}
\end{eqnarray}

\begin{eqnarray}
\frac{d \,\langle\hat{b}_{j,{\bm k}+{\bm k}'}^{\dagger}\hat{A}_{{\bm k},{\bm k}'}\rangle}{dt} &=&
-\Big(\gamma_j\big(1+\alpha_{\gamma}/V\sum_{{\bm k}''}\langle\hat{n}_{{\bm k}''}\rangle\big)
+\sum_{j',{\bm k}''}2\alpha_{\gamma}\gamma_{j'}/V\langle\hat{n}_{j',{\bm k}''}^{b}\rangle\Big)
\langle\hat{b}_{j,{\bm k}+{\bm k}'}^{\dagger}\hat{A}_{{\bm k},{\bm k}'}\rangle
\nonumber\\
&&
+i\frac{2g_j}{\sqrt{V}}|{\bm k}-{\bm k}'|\,
\Big(
\langle\hat{n}^{b}_{j,{\bm k}+{\bm k}'}\rangle
-2\langle\hat{n}_{{\bm k}}\hat{n}_{{\bm k}'}\rangle
\Big)
+i\frac{2g_j\alpha_g}{V^{3/2}}|{\bm k}-{\bm k}'|\,
\sum_{{\bm k}''}\Big(
\langle\hat{n}^{b}_{j,{\bm k}+{\bm k}'}\hat{n}_{{\bm k}''}\rangle
-2\langle\hat{n}_{{\bm k}}\hat{n}_{{\bm k}'}\hat{n}_{{\bm k}'',{\bm k}''}\rangle
\Big)\nonumber\\
\label{eq:baFieldNew}
\end{eqnarray}

\begin{eqnarray}
\frac{d \,\langle\hat{n}^{b}_{j,{\bm k}}\rangle}{dt} &=&
-2\gamma_j(1+\alpha_{\gamma}\sum_{{\bm k}'}\langle\hat{n}_{{\bm k}'}\rangle/V)\langle\hat{n}^{b}_{j,{\bm k}}\rangle
-\frac{2g_j}{\sqrt{V}}\sum_{{\bm k}'}
 |2{\bm k}'-{\bm k}| \,
 \Big(\text{Im}[\langle\hat{b}_{j,{\bm k}}^{\dagger}\hat{A}_{{\bm k}',{\bm k}-{\bm k}'}\rangle]
 +\alpha_g\sum_{{\bm k}''}\text{Im}[\langle\hat{b}_{j,{\bm k}}^{\dagger}\hat{A}_{{\bm k}',{\bm k}-{\bm k}'}\hat{n}_{{\bm k}''}\rangle]
 \Big),
\nonumber\\
 \label{eq:bbFieldNew}
\end{eqnarray}
where $\hat{n}^{b}_{j,{\bm k}}=\hat{b}_{j,{\bm k}}^{\dagger}\hat{b}_{j,{\bm k}}$ is the complex population operator.

Here we have neglected the kinetic energy term and the binding energy term. In addition, since the complex decay rate is large, it's fair to assume that the complex population can be neglected when it is compared to the molecule population. 
Therefore, by addiabatically eliminating the complex, one can obtain
\begin{eqnarray}
\langle\hat{b}_{j,{\bm k}+{\bm k}'}^{\dagger}\hat{A}_{{\bm k},{\bm k}'}\rangle&=&
-i\frac{4g_j}{\sqrt{V}}\big(1+\alpha_{g}n\big)|{\bm k}-{\bm k}'|
\langle\hat{n}_{{\bm k}}\rangle\langle\hat{n}_{{\bm k}'}\rangle
/\Big(\gamma_j\big(1+\alpha_{\gamma}n\big)\Big),
\\
\langle\hat{n}^{b}_{j,{\bm k}}\rangle&=&-\frac{g_j}{\sqrt{V}}\big(1+\alpha_{g}n\big)
\sum_{{\bm k}'} |2{\bm k}'-{\bm k}| \,
\text{Im}[\langle\hat{b}_{j,{\bm k}}^{\dagger}\hat{A}_{{\bm k}',{\bm k}-{\bm k}'}\rangle]
/\Big(\gamma_j\big(1+\alpha_{\gamma}n\big)\Big)\nonumber\\
&=&\frac{4g_j^2\big(1+\alpha_{g}n\big)^2}{V\gamma_j^2\big(1+\alpha_{\gamma}n\big)^2}
\sum_{{\bm k}'}|2{\bm k}'-{\bm k}|^2\langle\hat{n}_{{\bm k}-{\bm k}'}\rangle\langle\hat{n}_{{\bm k}'}\rangle,
\end{eqnarray}
where $n=\sum_{{\bm k}}\langle \hat{n}_{{\bm k}}\rangle/V$ is the  density of the molecules.

Similarly, by substituting the correlations into Eq.~(\ref{eq:ccFieldNew}), the dynamics for the molecular population becomes
\begin{eqnarray}
\frac{d \langle\hat{n}_{{\bm k}}\rangle}{dt}
&=&
-\sum_{j,{\bm k}'}\frac{16g_j^2\big(1+\alpha_{g}n\big)^2}{V\gamma_j\big(1+\alpha_{\gamma}n\big)}|{\bm k}-{\bm k}'|^2
\langle\hat{n}_{{\bm k}}\rangle\langle\hat{n}_{{\bm k}'}\rangle
\nonumber\\
&&-\sum_{j,{\bm k}',{\bm k}''}
\frac{8\alpha_{\gamma}g_j^2\big(1+\alpha_{g}n\big)}{V^2\gamma_j^2\big(1+\alpha_{\gamma}n\big)^2}|2{\bm k}''-{\bm k}'| 
\langle\hat{n}_{{\bm k}}\rangle\langle\hat{n}_{{\bm k}''}\rangle\langle\hat{n}_{{\bm k}'-{\bm k}''}\rangle\nonumber\\
&\approx&-\sum_{{\bm k}'}\Gamma_{{\bm k},{\bm k}'}\frac{\big(1+\alpha_{g}n\big)^2}{\big(1+\alpha_{\gamma}n\big)}
\langle\hat{n}_{{\bm k}}\rangle\langle\hat{n}_{{\bm k}'}\rangle
-\frac{1}{2}\sum_{j,{\bm k}',{\bm k}''}\frac{\alpha_{\gamma}}{V}
\Gamma_{{\bm k}'',{\bm k}'-{\bm k}''}
\langle\hat{n}_{{\bm k}}\rangle\langle\hat{n}_{{\bm k}''}\rangle\langle\hat{n}_{{\bm k}'-{\bm k}''}\rangle.
\label{eq:rateNew}
\end{eqnarray} 
Here we have assumed $\alpha_{g(\gamma)}  n\ll 1$  approximation that is found to be valid for the  KRb experimental parameters. In addition, we assume $\gamma_j\gg E_j$ since the only the close to resonance complex can be formed.

And the dynamical equation for the total number of the molecules is given by 
\begin{eqnarray}
\frac{d N}{dt}
&=&-\sum_{{\bm k},{\bm k}'}\Gamma_{{\bm k},{\bm k}'}
\big(1+2\alpha_{g}n-\alpha_{\gamma}n/2\big)
\langle\hat{n}_{{\bm k}}\rangle\langle\hat{n}_{{\bm k}'}\rangle
\end{eqnarray}
By comparing Eq.~(\ref{eq:rateNew}) with Eq.~(\ref{eq:rate}), we find the modified decay rate after taking into account of the inelastic molecule-complex collisions becomes 
\begin{eqnarray}
\Gamma_{{\bm k},{\bm k}'}^{P}=\Gamma_{{\bm k},{\bm k}'}\big(1+2\alpha_{g}n-\alpha_{\gamma}n/2\big),
\end{eqnarray}
indicating that the effective inelastic scattering parameter becomes
\begin{eqnarray}
g_{\text{im}}^{P}&=& g_{\text{im}}\big(1+2\alpha_{g}n-\alpha_{\gamma}n/2\big).
\end{eqnarray}
\section{Revised decay rates}
\label{App:decay_rate}

In a harmonic trap, the density of the gas is not homogeneous, therefore the spatial dependence of the effective scattering coefficient  $g_{\text{im}}^P({\bm r})$ needs to  be taken into account. 
This  leads to a revised decay rate $\Gamma_{{\bm n}_i {\bm n}_j {\bm n}_k {\bm n}_l}$ given by
\begin{eqnarray}
\Gamma_{{\bm n}{\bm n}'{\bm n}''{\bm n}'''}^{P}=\int d{\bm r}^{\,3} g^{P}_{\text{im}}(\bm r)
\big[\big({\bm \nabla}\phi^{*}_{{\bm n}}(\bm r)\big)\phi^{*}_{{\bm n}'}(\bm r)
-\phi^{*}_{{\bm n}}(\bm r)\big({\bm \nabla}\phi^{*}_{{\bm n}'}(\bm r)\big)\big]
\cdot\big[\big({\bm \nabla}\phi_{{\bm n}''}(\bm r)\big)\phi_{{\bm n}'''}(\bm r)
-\phi_{{\bm n}''}(\bm r)\big({\bm \nabla}\phi_{{\bm n}'''}(\bm r)\big)\big].
\nonumber\\
\label{eq:GammaIndex2}
\end{eqnarray}
Consequently, the revised rate equations for the mode populations are given by
\begin{eqnarray}
\frac{d \langle\hat{n}_{{\bm n}}\rangle}{dt}
&\approx-\sum_{{\bm n}'}\Gamma_{{\bm n}{\bm n}'}\langle\hat{n}_{{\bm n}}\rangle\langle\hat{n}_{{\bm n}'}\rangle,
\label{eq:rateHarmonic2}
\end{eqnarray}
where $\Gamma_{{\bm n}{\bm n}'}^{P}\equiv4\Gamma_{{\bm n}{\bm n}'{\bm n}{\bm n}'}^{P}$

The scaling of $\overline{\Gamma}^P$ for systems with a large number of particles is limited by the computation complexity.
To overcome this limit, here we instead take the local density approximation starting from a semi-classical phase space distribution given by 
\begin{eqnarray}
f({\bm r},{\bm p})&=&\frac{1}{\exp[(\frac{m\omega^2 {\bm r}^2}{2}+\frac{{\bm p}^2}{2m}-\mu)/k_BT]+1},
\end{eqnarray}
the averaged decay rate can be calculated as
\begin{eqnarray}
\overline{\Gamma}^P&=&\frac{\int\, g_{\text{im}}^P({\bm r})\,{\bm p}^2\, f({\bm r},{\bm p})\, d{\bm r}^3 d{\bm p}^3}
{NV}
\label{eq:light-induce-2}
\end{eqnarray}
where $N$ and $V$ denotes the particle number and the volume respectively, 
and the first term in the integrand $g_{\text{im}}^P({\bm r})$ accommodates the spatial dependence,
the second term ${\bm p}^2$ represents the $p$-wave collisional kernel that is proportional to the kinetic energy of the gas,
and the denominator is simply the total particle number of the system.

\begin{figure}[hbt!]
\centering
\includegraphics[width=0.85\textwidth]{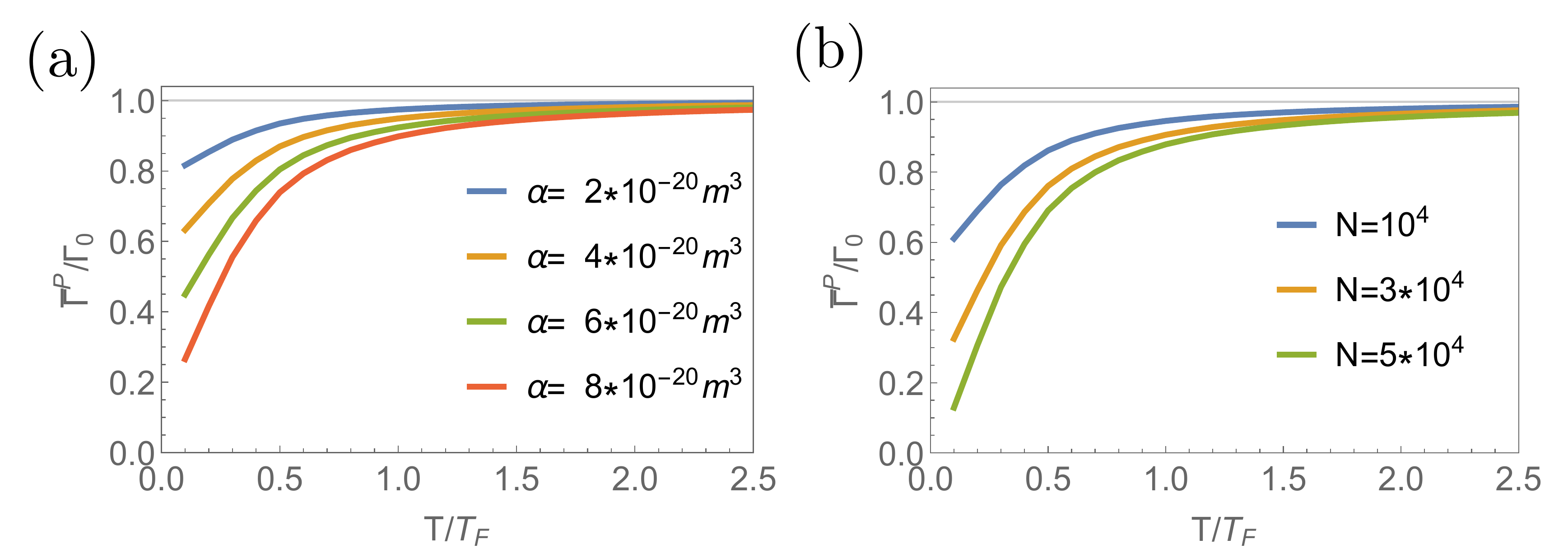}
\caption{The plots show $\overline{\Gamma}_P/\Gamma_0$  for  different (a) $\alpha$ values assuming $N=2\times 10^4$ and (b) different particle number $N$ setting  $\alpha=6\times 10^{-20}\text{m}^3$.}
\label{fig:modified-decay}
\end{figure}

We compute the integral Eq.~(\ref{eq:light-induce-2}) numerically assuming different $\alpha$ and particle number $N$.
As shown in Fig.~\ref{fig:modified-decay}, we find that the ratio $\overline{\Gamma}_P/\Gamma_0$  assuming $\alpha =0$)
saturates at high temperature and gets suppressed as the gas enters quantum degeneracy ($\Gamma_0$ is calculated using Eq.~(\ref{eq:light-induce-2}). 
In addition, the degree of suppression and the saturation temperature increase with increasing  particle number. 

\section{Fitting Analysis}
\label{App:compare_exp}
In the experiment, the molecules are created and cooled down to the Fermi degenerate regime.
By fitting the initial density profile to a Fermi-Dirac distribution, the initial temperatures $T_0^{\text{ex}}$ and $T_0^{\text{ex}}/T_F$ are obtained.
To keep track of the reactive collision processes, 
the particle number $N^{\text{ex}}(t)$ and the volume $V^{\text{ex}}(t)$ are measured as a function of the evolution time $t$. 
To compare with the experimentally extracted  decay rate, both 
the experimental initial energy density and the initial particle number are needed as an input parameters for the  theory. They are extracted by a fitting procedure: 
By fitting $V^{\text{ex}}(t)$ to $V(t)=(4\pi\epsilon(t)/3m\overline{\omega}^2)^{3/2}$, 
the initial energy density $\epsilon_0^{\text{ex}}\pm\Delta\epsilon^{\text{ex}}$ 
and the linear heating rate $h^{\text{ex}}\pm \Delta h^{\text{ex}}$ can be extracted 
with $\Delta\epsilon^{\text{ex}}$ and $\Delta h^{\text{ex}}$ the uncertainties.
Furthermore, by finding the best fit of $N^{\text{ex}}(t)$ to the theoretical $N^{\text{th}}(t)$ obtained, the initial particle number $N_0^{\text{ex}}\pm\Delta N_0^{\text{ex}}$ can be obtained. 

\begin{figure}[hbt!]
\centering
\includegraphics[width=0.35\textwidth]{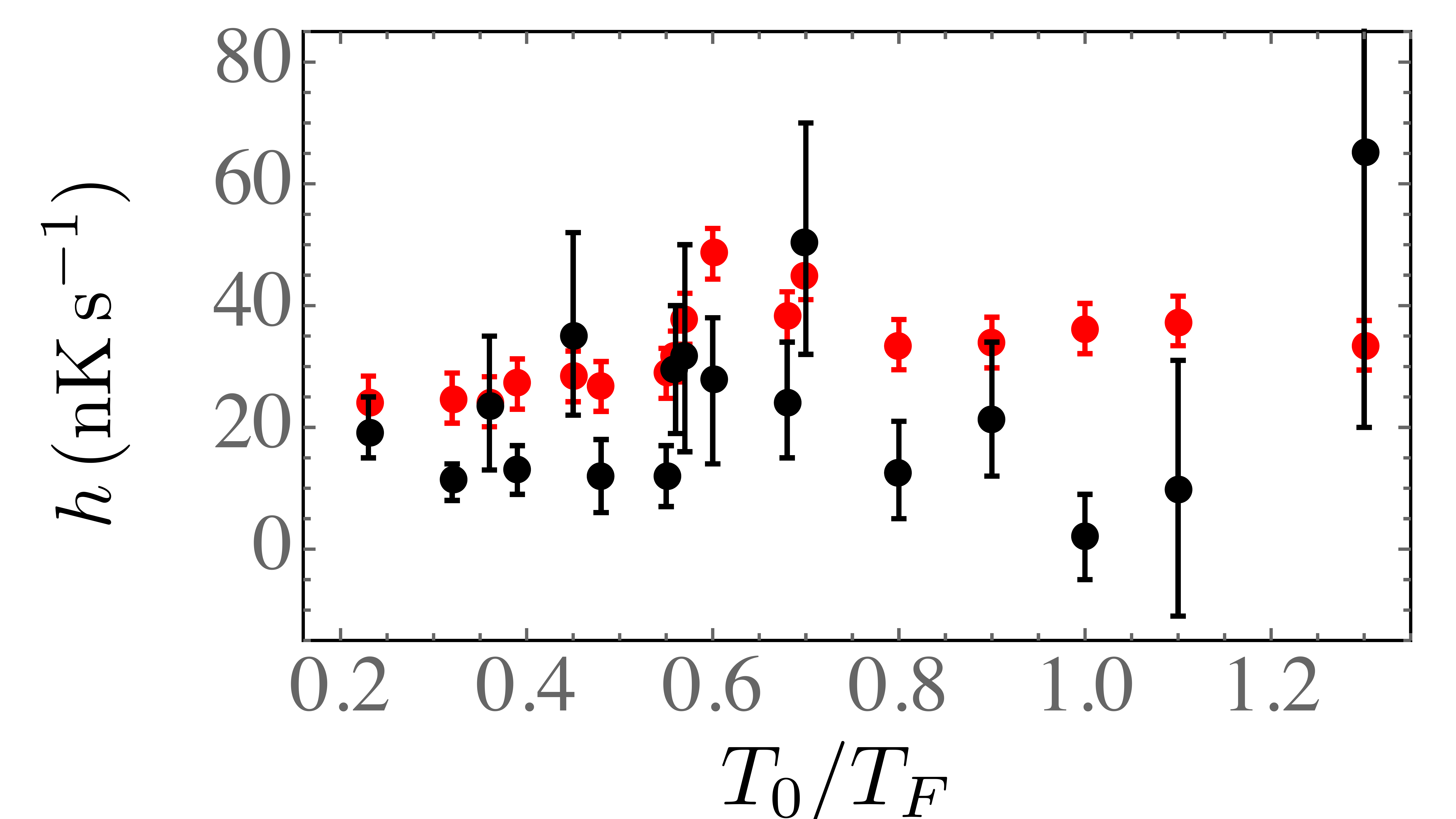}
\caption{  Comparison between the theoretically   predicted (red dots) and the experimental measured (black dots)  heating rates. }
\label{fig:Heating}
\end{figure}

Assuming $\alpha_0 = 0.07$ and $h_{\text{bg}} = 20\pm 4\text{nK/s}$, 
together with the extracted  parameters $N_0^{\text{ex}}$, $\epsilon_0^{\text{ex}}$ can be solved self-consistently.
The theoretically predicted $h^{\text{th}}$ is extracted from a linear fit  to  $\epsilon^{\text{th}}(t)$. 
The comparison of the theoretically  predicted $h^{\text{th}}$ and the experimentally  measured $h^{\text{ex}}$ are shown in Fig.~\ref{fig:Heating}.
We find that  for the fixed $h_{\text{bg}}$ used in the theory model, the theory  results roughly agree with the experimental ones  in the degenerate regime where the density is high,
while the theory overestimates the heating rates in the classical regime where the density is low, which is qualitatively consistent with the conjecture that the background heating is induced by the density-dependent collisions and should be smaller for dilute systems.

\begin{figure}[hbt!]
\centering
\includegraphics[width=0.95\textwidth]{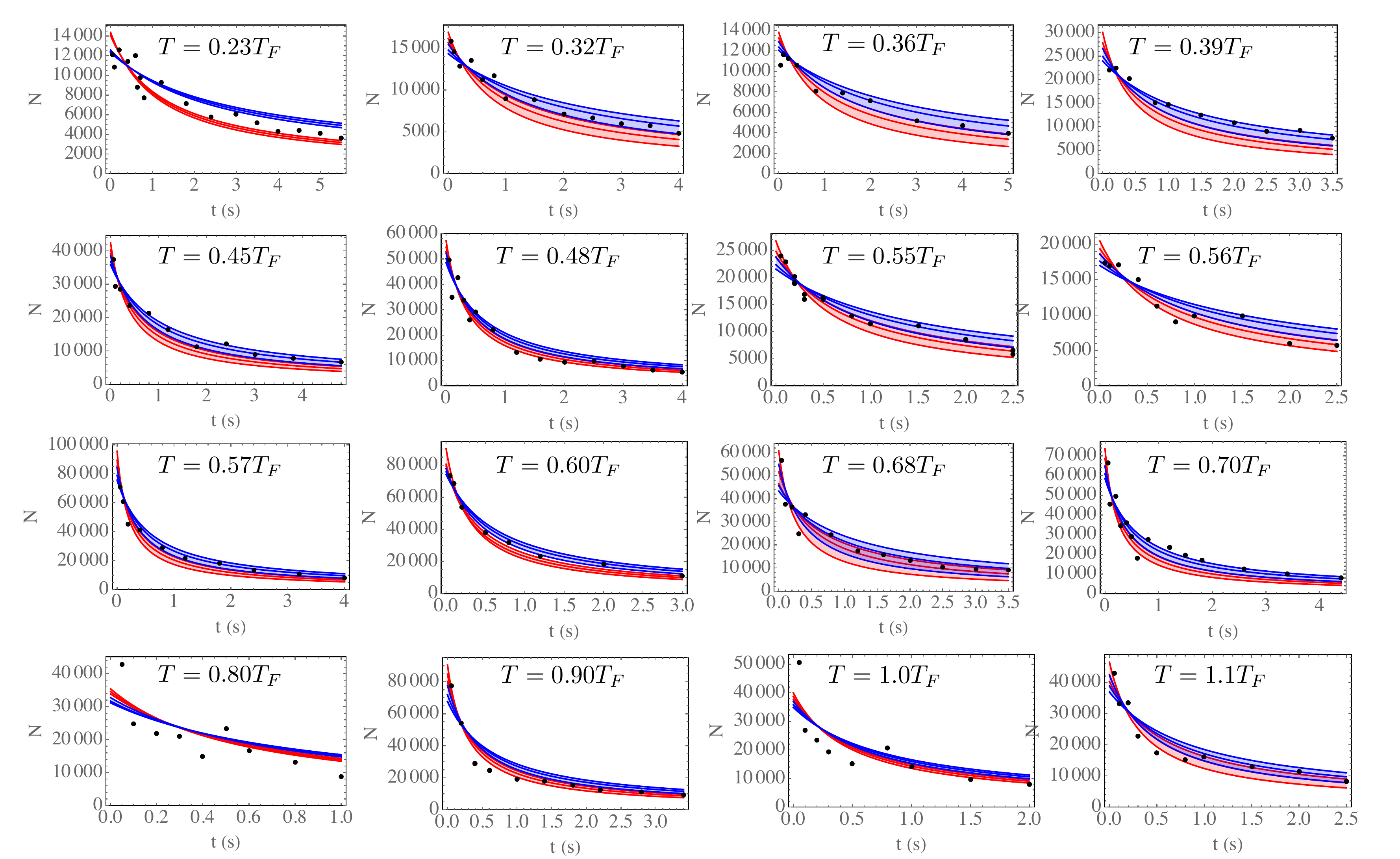}
\caption{
Comparison between theoretical results and experimental data (black dots) for the particle number $N(t)$.
The theoretical results are obtained using the decay rates (red bands) $\overline{\Gamma}_0^{\text{th}}$ and the revised decay rates (blue bands) $\overline{\Gamma}_0^{\text{th},P}=\overline{\Gamma}_0^{\text{th}}\times (\overline{\Gamma}^P/\Gamma_0)$ respectively, where the ratio $\overline{\Gamma}^P/\Gamma_0$ is calculated assuming $\alpha=8\times 10^{-20}\text{m}^3$.
}
\label{fig:comparison_plot}
\end{figure}

The theory predicted $N(t)$ is obtained by substituting theoretically calculated decay rate $\overline{\Gamma}_0^{\text{th}}$, the heating rate $h^{\text{th}}$, and the experimentally measured initial conditions $\epsilon_0^{\text{ex}}$ and $N_0^{\text{ex}}$
into Eq.~(\ref{eq:solution_theory}).
In Fig.~\ref{fig:comparison_plot}, we  compare  the dynamics of  $N(t)$ predicted by  the theoretical results and the experimental data. 
The decay rate $\beta_0$ is obtained as the best fit of the theoretical $n(t)=N(t)/V(t)$ to Eq.~(\ref{eq:solution_theory}).

To incorporate the effect of the formation of the complex, we replace $\overline{\Gamma}_0^{\text{th}}$ by  $\overline{\Gamma}_0^{\text{th},P}=\overline{\Gamma}_0^{\text{th}}\times (\overline{\Gamma}^P/\overline{\Gamma}_0)$. Since $\overline{\Gamma}^P/\overline{\Gamma}_0$ gives rise to suppression, 
the agreement between the dynamics of the particle number of the theoretical results and experimental data becomes better, as shown in Fig.~\ref{fig:comparison_plot}.

\end{document}